\documentclass[doublecol]{epl2}
\usepackage{mathrsfs,stfloats}
\usepackage{amssymb}
\usepackage{amsfonts}
\usepackage{amsmath}
\usepackage{latexsym}
\usepackage{graphicx}

\title{Multiple Hybrid Phase Transition: Bootstrap Percolation on Complex Networks with Communities}
\shorttitle{Bootstrap Percolation on Complex Networks with community structure} 

\author{Chong Wu\inst{1}, Shenggong Ji\inst{1}, Rui Zhang\inst{1,2,3}, Liujun Chen\inst{4}\footnote{E-mail: chenlj@bnu.edu.cn}, Jiawei Chen\inst{4}, Xiaobin Li\inst{1} \and Yanqing Hu\inst{1}\footnote{E-mail: yanqing.hu.sc@gmail.com}}
\shortauthor{Chong Wu\etal}

\institute{
 \inst{1} School of Mathematics, Southwest Jiaotong University, Chengdu  610031, China\\
 \inst{2} Levich Institute and Physics Department, City College of the City University of New York, New York, NY 10031, USA\\
 \inst{3} Institute for Molecular Engineering, University of Chicago, Chicago, Illinois 60637, USA\\
 \inst{4} School of Systems Science, Beijing Normal University, Beijing 100875, China}

\abstract{Bootstrap percolation is a well-known model to study the spreading of rumors, new products or innovations on social networks. The empirical studies show that community structure is ubiquitous among various social networks. Thus, studying the bootstrap percolation on the complex networks with communities can bring us new and important insights of the spreading dynamics on social networks. It attracts a lot of scientists' attentions recently. In this letter, we study the bootstrap percolation on Erd\H{o}s-R\'{e}nyi networks with communities and observed second order, hybrid (both second and first order) and multiple hybrid phase transitions, which is rare in natural system. Moreover, we have analytically solved this system and obtained the phase diagram, which is further justified well by the corresponding simulations.}

\begin{document}
\maketitle

\section{\label{sec1}Introduction}
Bootstrap percolation is essentially an activation process via connectivity links on networks. It adopts as follows: each node is either active (at state 1) or inactive (at state 0). Initially, a fraction of nodes are set up at state 1 and the rest nodes remain at 0. At each step, the node at state 0 will change its state to 1 permanently if at least $k$ of its neighbors are in state 1. This process are extensively applied to model the information propagation among individuals \cite{Watts02,epidemic1,epidemic2,epidemic3,general6,hu_prl}. It has been studied on d-dimensional lattices \cite{H2,BB,H1,CC}, infinite trees \cite{BPP}, random regular graph \cite{BP,FS} and even finite random graph \cite{pre16}. Recently it is successfully applied to unstructured Erd\H{o}s-R\'{e}nyi(ER) networks \cite{BDGM10}.

Many empirical studies show the social network has communities\cite{GN, GMT}, which are the functional groups and plays a crucial role in the network\textcolor{black}{\cite{HKBB,RB04, RB07, SKK2014, GCD, KLN, HNYCFD}}. Obviously, it is worth noting that by various reasons to study the bootstrap percolation on complex networks with communities. Bi-community structure (only two communities in a network) is simple and representative, as multi-community system can reduce to bi-community one in the ¡°one vs all¡± manner\cite{HDFD}. Thus, in this letter we focus on bi-community structure networks and all the equations can be generalized to multi-community system.

To model, each community is assigned a different probability $f_i$($i=1,2$) by which the nodes are randomly chosen to be initially active, otherwise inactive in the beginning. Next we exam all inactive nodes. If any of them has at least $k$ active neighbors, it becomes activated. This activation, by increasing the number of active nodes, may potentially make more inactive nodes to become activated. Therefore we repeat this process until no more nodes can be activated, namely we reach a final state.

We analytically express the fractions of active nodes and also the giant component size in each community in the final state as functions of the initial fractions of active nodes. We find that such functions undergo multiple hybrid\cite{CC,HKCH} transitions, and a discontinuous jump of a function for one community may trigger a simultaneous jump of the function for the other community. We have further obtained the phase diagram of the total number of jumps in terms of the inner-degrees of the two communities for ER networks. We also discuss the emergence of such hybrid phase transitions.

\section{\label{sec2}Analytical Model}
We consider a single complex network with community structure. A community is a subnetwork of the network, and the connections between the nodes in the community are much closer than the connections between node in the community and node out of the community. Bi-community structure is representative as it can be easily generalized to multi-community. Therefore we mainly study the bootstrap percolation on an undirected network with two communities, namely community 1 and 2. Let $P^1(i,j)$ and $P^2(i,j)$ be the probability distribution of inner-degree and outer-degree of a node in community 1 and 2 respectively, with $i$ the number of edges of a node connected to community 1, and $j$ the number of edges of a node connected to community 2. In $P^1(i,j)$, $i$ is the inner-degree and $j$ is the outer-degree, whereas in $P^2(i,j)$, $i$ is the outer-degree and $j$ is the inner-degree.
Let $f_i$ be the probability of a node in community $i(i=1,2)$ being initially active. An initially inactive node becomes active if it has at least $k$ active downstream neighbors in either community 1 or 2. These active neighbors must satisfy the same criteria, that they are either active in the very beginning or they have $k$ further downstream neighbors of their own that are previously active. A final (equilibrium) state is reached when no more nodes can be activated by the above criteria.

At equilibrium of the bootstrap percolation process, a fraction $S_i$ (with $i=1,2$) of nodes in community $i$ are active. $S_i$ can also be interpreted as the probability that an arbitrarily chosen node in community $i$ is active in the final state. In the following we are also interested in calculating the probability of an arbitrarily chosen node in each community belonging to the giant active component of the whole network, denoted by $S_{gc1}$ and $S_{gc2}$ respectively.
To calculate $S_{1}$ and $S_{2}$, we randomly choose an edge from community $i$ to community $j(i,j=1,2)$, and let $Z_{i,j}$ denote the probability that the node arrived at is active in the final state. In other words, the node is either active in the beginning, or has at least $k$ active downstream neighbors reached by edges except the one we arrive from.

Randomly choose an edge in community 1, then there are two possibilities for the arriving node to be active: one is that the node is active in the beginning, which has a probability $f_1$; the other is that the node is inactive initially, but it has $k$ active downstream neighbors. These neighbors must also be in one of the two possibilities to be active. So if the node we arrived at by the chosen edge has $i$ other edges in community 1 except the one we arrived from and $j$ neighbors in community 2, then the probability is $\frac{(i+1)P^1(i+1,j)}{\sum_{i\geq 0, j\geq 0}iP^1(i,j)}$. If among the $i$ edges in community 1, there are $l$ edges connected to active nodes in community 1, and among the $j$ edges linked with community 2, there are $m$ edges connected to active nodes, then $l$ and $m$ should satisfy $l+m\geq k$. So we can construct the following equation for $Z_{11}$:

\begin{align*}
Z_{11}&=f_1+(1-f_1)\sum_{i+j\geq k}\frac{(i+1)P^1(i+1,j)}{\sum_{i\geq 0,j\geq 0}iP^1(i,j)}
\sum_{\begin{subarray}{l}l+m=k\\ l\leq i,m\leq j\end{subarray}}^{i+j}\nonumber\\
& \binom{i}{l}Z_{11}^l(1-Z_{11})^{i-l}\binom{j}{m}Z_{12}^m(1-Z_{12})^{j-m}.
\end{align*}
Similarly, we can obtain the equations for $Z_{12}$, $Z_{21}$ and $Z_{22}$.

A randomly chosen node in community $i$ is active at equilibrium if either it is active at first, or it has at least $k$ edges connected with active downstream neighbors in either community 1 or 2. So by writing down mathematical expressions for the probabilities, we can construct the following equation for $S_{1}$:

\begin{align*}
S_{1}&=f_1+(1-f_1)\sum_{i+j\geq k}P^1(i,j)
\sum_{\begin{subarray}{l}l+m=k\\l\leq i,m\leq j\end{subarray}}^{i+j}\nonumber\\
& \binom{i}{l}Z_{11}^l(1-Z_{11})^{i-l}\binom{j}{m}Z^m_{12}(1-Z_{12})^{j-m}.
\end{align*}
We can also get the equation for $S_2$, and these equations can be solved numerically for a given network degree distribution.

\section{Results}
\subsection{\quad Erd\H{o}s-R\'{e}nyi Networks}
The Erd\H{o}s-R\'{e}nyi (ER) network (which has a Poisson degree distribution in the infinite size limit) is a representative random graph, so we discuss ER network in particular, although our equations can be applied to any random networks. In ER network, the presence or absence of an edge between two nodes is independent of the presence or absence of any other edge, so that each edge may be considered to be present with independent probability $p$. The degree of any particular node (that is the number of edges of the node) has a probability distribution $p_k$ given by $p_k=\binom{N}{k}p^k(1-p)^{N-k}\approx\frac{z^k e^{-z}}{k!}$, where $N$ is the number of nodes in the network, $z$ is the average degree of any node, and the second equality becomes exact in the limit of large $N$\cite{BDGM10}.

In the ER network consisting of two communities, suppose any edge from a node in community $i$ to a node in community $j$ is present with probability $p_{ij}(i,j=1,2)$. So we have
\begin{align}
P^1(i,j)&=\binom{N_1}{i}p_{11}^i(1-p_{11})^{N_1-i}\binom{N_2}{j}p_{12}^j(1-p_{12})^{N_2-j}\nonumber\\
&\approx\frac{k_{11}^i e^{-k_{11}}}{i!}\cdot\frac{k_{12}^j e^{-k_{12}}}{j!},\nonumber
\end{align}

where $k_{ij}$ is the average edges of any node in community $i$ connecting with nodes in community $j$, \textcolor{black}{$N_i$ is the number of nodes in community $i$}. In ER network, we get that
\begin{align*}
Z_{11}&=Z_{21}=S_{1},\\
Z_{22}&=Z_{12}=S_{2},
\end{align*}
and
\begin{equation}\label{S_1}
S_1=f_1+(1-f_1)\sum_{r\geq k}\frac{(S_1k_{11}+S_2k_{12})^r}{r!} e^{-(S_1k_{11}+S_2k_{12})},
\end{equation}
\begin{equation}\label{S_2}
S_2=f_2+(1-f_2)\sum_{r\geq k}\frac{(S_1k_{21}+S_2k_{22})^r}{r!} e^{-(S_1k_{21}+S_2k_{22})},
\end{equation}
where $k_{11}=N_1\cdot p_{11}$, $k_{12}=N_2\cdot p_{12}$, $k_{21}=N_1\cdot p_{21}$, $k_{22}=N_2\cdot p_{22}$, and $p_{12}=p_{21}$.

\begin{figure}
\centering
\includegraphics[scale=0.5]{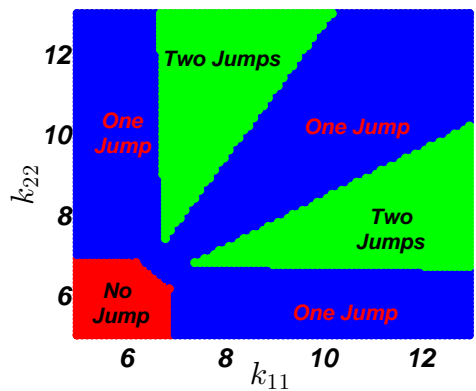}
\caption{The diagram of the phase transitions \textcolor{black}{by theoretical calculations}. Given values of $k_{12}$, $k_{21}$ and $k$, the function $S(f)$(also $S_i(f)$) depends on the values of $k_{11}$ and $k_{22}$. Here $k_{12}=k_{21}=0.5$, $k=5$. \textcolor{black}{If the value of $S(f_0+0.001)-S(f_0)$ is larger than 0.05, we think $f_0$ is a jump point of $S(f)$.} Given the values of $k_{11}$ and $k_{22}$ in different areas in the above graph, there may be one or two discontinuous jumps, or no jumps in the size of the active component of the network as a function of initial active fraction $f$. }
\label{fig1}
\end{figure}

Suppose communities 1 and 2 are of identical size, and their initial fractions of active nodes are identical. Given values of $k_{ij}$, the size of the active component in the equilibrium is a function of the initial fraction of active nodes $f$.
 \textcolor{black}{ In order to show the consistence between theoretical results and simulations, we show in appendix in \cite{WJZCCLH} the active fraction $S=(S_1+S_2)/2 $ (when $N_1=N_2$) as a function of the initial active fraction $f_1=f_2=f$ for randomly chosen groups of values of
$k_{ij} $ in an ER network, both theoretically and by simulations.}
 We find that there are three subcases. There may be one or two discontinuous jumps, or no jumps in curve of the size of the active component as a function of $f$. Fig.~\ref{fig1} shows the diagram of the phase transitions, with respect to \textcolor{black}{given values of $k_{12}=k_{21}=0.5$}. \textcolor{black}{The phase diagram depends on the inner-degrees and outer-degrees of the network $k_{ij}$. When $N_1\neq N_2$, $k_{12}$ will not be equal to $k_{21}$ either, then the diagram will be different from Fig.~1. We showed different phase diagrams when $N_1=N_2$, $2N_1=N_2$, $3N_1=N_2$ respectively, and these figures can be seen  in the appendix in \cite{WJZCCLH}. }
Given small values of $k_{ij}$, for example $k_{11}=6.5$ and $k_{22}=6.5$, the jump phenomenon does not appear. Once there is a jump in $S_1$ or $S_2$, then the jump appears in $S$. When $k_{11}=10$, $k_{22}=7$, there are two jumps in $S$, as $S_1$ and $S_2$ jump at two different values of $f$.
In the network with communities, we find a phenomenon which is different from single network without communities. As function of $f$, $S_1$ and $S_2$ interact with each other, so if one of them has a change, the other one will change immediately. Their interactions depend on the interconnection of the two communities. The larger the value of $k_{12}$ or $k_{21}$ is, the stronger the interactions between $S_1$ and $S_2$ are.

The equations above can be solved numerically. If multiple solutions exist, the smallest value is always the physical solution. To find the location of the discontinuous jump of $S_i$, we can observe that the jump appears at the disappearance of the smallest solution of Eq.~\ref{S_1} and Eq.~\ref{S_2}.

\begin{figure*}
\centering
\includegraphics[scale=0.38]{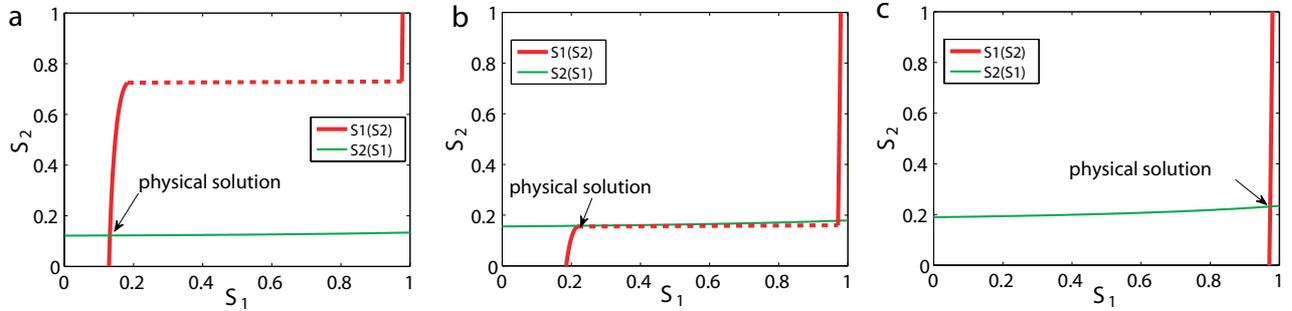}
\caption{(Color online)The process of the jump of the physical solution. (a) The curves of the two functions $S_1(S_2)$ and $S_2(S_1)$ intersecting at the smallest solution(physical solution). Here $f=0.12$, $k_{11}=10$, $k_{22}=7$, $k_{12}=0.5$, $k_{21}=0.5$, $k=5$. (b) The curves intersecting at the smallest solution, which is also the jump point of function $S_1(S_2)$(see figure 4). Here $f=f_{c1}\approx0.15$, $k_{11}=10$, $k_{22}=7$, $k_{12}=0.5$, $k_{21}=0.5$, $k=5$. (c) The disappearance of the smallest solution. Here $f=0.18$, $k_{11}=10$, $k_{22}=7$, $k_{12}=0.5$, $k_{21}=0.5$, $k=5$.}
\label{fig2}
\end{figure*}

Fig.~\ref{fig2} shows the process of the disappearance of the smallest solution, which explains the jump in $S_i(f)$. To find the location of the discontinuous transition theoretically, we argue in the following way. Given the values of $k_{ij}$, $f_1=f_2=f$ and $k$, Eq.~\ref{S_1} determines $S_1$ as a function of $S_2$, i.e. $S_1(S_2)$, and Eq.~\ref{S_2} determines $S_2$ as a function of $S_1$, i.e. $S_2(S_1)$. The intersection point of the curves of the two functions is the solution. We notice that the functions $S_1(S_2)$ and $S_2(S_1)$ may be discontinuous, as shown in Fig.~\ref{fig2}. Given $k_{11}=10$, $k_{22}=7$, $k_{12}=0.5$, $k_{21}=0.5$, $k=5$, there are two jumps in the curve of function $S(f)$, also in the curves of functions $S_i(f)$. The jump points are $f=f_{c1}\approx0.15$ and $f=f_{c2}\approx0.22$. When $f=f_{c1}\approx0.15$, the function $S(f)$(also $S_i(f)$) jumps . Meanwhile, in Fig.~\ref{fig2} (b), the physical solution point is the jump point of the curve of function $S_1(S_2)$. Therefore, we can see that at the jump point $f_c$ of function $S(f)$ (also $S_i(f)$), the smallest solution of Eq.~\ref{S_1} and Eq.~\ref{S_2} is related to the jump point of curve of one of the two functions $S_1(S_2)$ and $S_2(S_1)$.

The jump point of $S_1(S_2)$ can be found in the following way. Given values of $k_{ij}$ and $k$, let $f=f_{c1}$, and let $S_2=\tilde{S_2}$, which is the value of $S_2$ of the physical solution. Let $F(S_1)$ denote the right side of Eq.~\ref{S_1} as a function of $S_1$, and let $G(S_1)=S_1$. Then the value of $S_1$ at the intersection point of curves of the two functions $F(S_1)$ and $G(S_1)$ is the value of $S_1$ of the physical solution, denoted by $\tilde{S_1}$ as before.

\begin{figure*}
\centering
\includegraphics[scale=0.38]{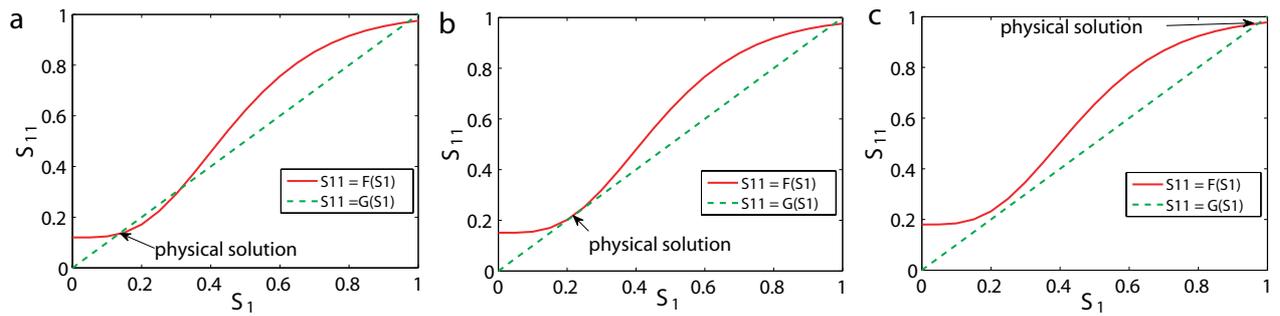}
\caption{(Color online)The appearance of the jump point of function $S_1(S_2)$, which is also the process of the disappearance of the smallest solution of $F(S_1)=G(S_1)$. (a) The curves of the two functions $G(S_1)$ and $F(S_1)$ intersecting at the physical solution. Here $f=0.12$, $k_{11}=10$, $k_{22}=7$, $k_{12}=0.5$, $k_{21}=0.5$, $k=5$. (b) The curves of the two functions $G(S_1)$ and $F(S_1)$ being tangential to each other at the smallest intersection. Here $f=f_{c1}\approx0.15$, $k_{11}=10$, $k_{22}=7$, $k_{12}=0.5$, $k_{21}=0.5$, $k=5$. (c) The disappearance of the smallest solution of $F(S_1)=G(S_1)$. Here $f=0.18$, $k_{11}=10$, $k_{22}=7$, $k_{12}=0.5$, $k_{21}=0.5$, $k=5$.}
\label{fig3}
\end{figure*}

Fig.~\ref{fig3} shows the appearance of the jump point $f=f_{c1}\approx0.15$ of function $S_1(S_2)$, which is also the process of the disappearance of the smallest solution of $F(S_1)=G(S_1)$.
To be precise, we arrive at the jump point $f_{c1}\approx0.15$ of $S_1(S_2)$

\begin{equation}\label{eq5}
\frac{\mathrm{d}F(S_1)}{\mathrm{d}S_1}=1.
\end{equation}
Similarly, we can analyze the appearance of the jump point $f=f_{c2}\approx0.22$ of $S_2(S_1)$. The jump point $f_{c2}\approx0.22$ of $S_2(S_1)$ can be found by
\cite{BPPSH10}

\begin{equation}\label{eq6}
\frac{\mathrm{d}H(S_2)}{\mathrm{d}S_2}=1,
\end{equation}
where $H(S_2)$ denotes the right side of Eq.~\ref{S_2}, with given values of $k_{ij}$ and $k$, and $S_1=\tilde{S}_1$, $f=f_{c2}$. Finally, the jump points $f_c$ of functions $S_i(f)$ can be found from any solution of one of the two groups of equations: Eq.~\ref{S_1}, Eq.~\ref{S_2} and Eq.~\ref{eq5}, or Eq.~\ref{S_1}, Eq.~\ref{S_2} and Eq.~\ref{eq6}. If there are two different solutions, then there are two jumps in function $S(f)$.
\subsection{\quad The analysis of the fraction of the giant active component $S_{gci}$}
Now we consider the probability $S_{gci}(i=1,2)$ that an arbitrarily chosen node in community $i$ belongs to the giant active component \textcolor{black}{of the whole network}. In the infinite size limit, the giant active component is an active subtree of infinite extent. Define $X_{ij}$ to be the probability that the node arrived at by following an arbitrarily chosen edge from community $i$ to community $j$ satisfying the conditions for $Z_{ij}$ and has at least one edge leading to an active subtree of infinite extent except the arbitrarily chosen edge. There are two possibilities: one is that the node arrived at is active at first and has at least one edge leading to an active subtree of infinite extent; the other one is that the node arrived at is inactive initially, but it has more than $k$ edges, except the arbitrarily chosen edge, leading to active nodes at equilibrium in community 1 or 2, and at least one of those edges leads to an active subtree of infinite extent.

Writing down the mathematical expressions of the probabilities for the above two possibilities, we get

\begin{align*}
X_{11}&=f_1\sum_{i+j\geq 1}\frac{(i+1)P^1(i+1,j)}{\sum_{i\geq 0,j\geq 0}iP^1(i,j)}\biggl[\sum_{\begin{subarray}{l}0\leq m\leq i\\0\leq n\leq j\\m+n\geq 1\end{subarray}}\nonumber\\
&\binom{i}{m}X_{11}^m(1-X_{11})^{i-m}\binom{j}{n}X_{12}^n(1-X_{12})^{j-n}\biggr]\nonumber\\
&+(1-f_1)\sum_{i+j\geq k}\frac{(i+1)P^1(i+1,j)}{\sum_{i\geq 0,j\geq 0}iP^1(i,j)}\biggl\{\sum_{\begin{subarray}{l}k\leq m+n\leq i+j\\
0\leq m\leq i\\0\leq n\leq j\end{subarray}}\nonumber\\&\binom{i}{m}\binom{j}{n}
\biggl[\sum_{\begin{subarray}{l}1\leq s+t\leq m+n\\0\leq s\leq m\\0\leq t\leq n \end{subarray}}\binom{m}{s}\binom{n}{t}X_{11}^s(Z_{11}-X_{11})^{m-s}\nonumber\\
& X_{12}^t(Z_{12}-X_{12})^{n-t}(1-Z_{11})^{i-m}(1-Z_{12})^{j-n}\biggr]\biggr\}
\end{align*}
\textcolor{black}{The details about this equation can be seen in appendix C in \cite{WJZCCLH}.}

The equations of $X_{12}$, $X_{21}$ and $X_{22}$ are similar. And we get the probability $S_{gc1}$

\begin{align}\label{S {gc1}}
S_{gc1}&=f_1\sum_{i+j\geq 1}P^1(i,j)\biggl[\sum_{\begin{subarray}{l}0\leq m\leq i\\ 0\leq n\leq j\\m+n\geq 1\end{subarray}}\binom{i}{m} \binom{j}{n} \nonumber\\
& X_{11}^m (1-X_{11})^{i-m} X_{12}^n (1-X_{12})^{j-n}\biggr] \nonumber\\
& +(1-f_1)\sum_{i+j\geq k}P^1(i,j)\biggl\{\sum_{\begin{subarray}{l}k\leq m+n\leq i+j\\ 0\leq m\leq i\\0\leq n\leq j\end{subarray}}\binom{i}{m}\binom{j}{n}\nonumber\\
& \biggl[\sum_{\begin{subarray}{l}1\leq s+t\leq m+n\\0\leq s\leq m\\0\leq t\leq n \end{subarray}}\binom{m}{s}\binom{n}{t}X_{11}^s(Z_{11}-X_{11})^{m-s}\nonumber\\
& X_{12}^t(Z_{12}-X_{12})^{n-t}(1-Z_{11})^{i-m}(1-Z_{12})^{j-n}\biggr]\biggr\}
\end{align}
The equation for $S_{gc2}$ can be obtained similarly.

For ER network, $$\frac{(i+1)P^1(i+1,j)}{\sum_{i\geq 0,j\geq 0}iP^1(i,j)}=P^1(i,j)$$, so
\begin{align*}
X_{11}&=X_{21}=S_{gc1}, \\
X_{12}&=X_{22}=S_{gc2}.
\end{align*}

Finally, \textcolor{black}{substituting $S_{gc1}$  and  $S_{gc2}$ for $X_{11}$ and $X_{12}$ respectively on the right side of Eq.~\ref{S {gc1}}, we get the final equation for $S_{gc1}$.}
The equation of $S_{gc2}$ is just like $S_{gc1}$ in ER network.

\begin{figure}
\centering
\includegraphics[scale=0.45]{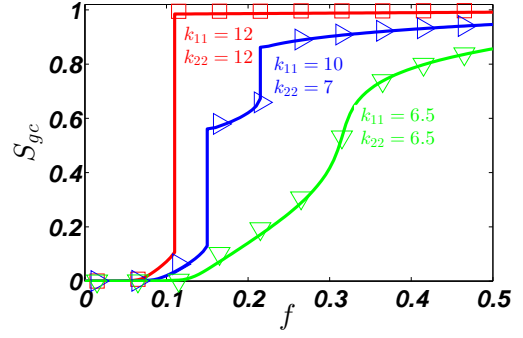}
\caption{(Color online) The fraction $S_{gc}$ of the giant active component in the whole network, for both the theory and simulations. Symbols are simulation data. Solid lines are theoretical solutions. We suppose $S_{gc}=\frac{S_{gc1}+S_{gc2}}{2}$, and $f_1=f_2=f$. Here we set $k_{12}=0.5$, $k_{21}=0.5$, and $k=5$. When $k_{11}=12$, $k_{22}=12$, there is one jump(the red line). When $k_{11}=10$, $k_{22}=7$, there are two jumps(the blue line). When $k_{11}=6.5$, $k_{22}=6.5$, there is no jump(the green line). }
\label{fig4}
\end{figure}

\begin{figure}
\centering
\includegraphics[scale=0.45]{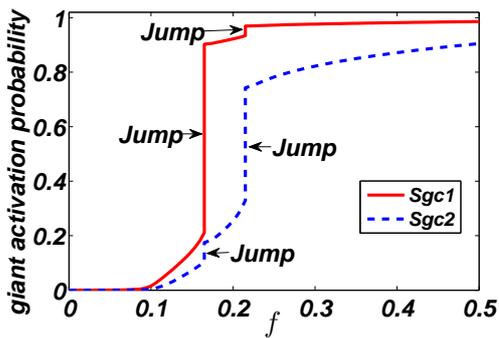}
\caption{(Color online) The interaction between $S_{gc1}$ and $S_{gc2}$. There are two discontinuous jumps in both $S_{gc1}$ and $S_{gc2}$. When a jump appears in one of $S_{gci}(i=1,2)$, a jump appears in the other one too. Here $k_{11}=8$, $k_{22}=7$, $k_{12}=2.4$, $k_{21}=0.5$, $k=5$.}
\label{fig5}
\end{figure}

Fig.~\ref{fig4} shows the giant active component fraction $S_{gc}=(S_{gc1}+S_{gc2})/2$ \textcolor{black}{(when $N_1=N_2$)} in the whole network as a function of the initial active fraction $f_1=f_2=f$ for three groups of values of $k_{ij}$ in an ER network, both theoretically and by simulations. \textcolor{black}{If $N_1\neq N_2$, then $S_{gc}=N_1/(N_1+N_2)S_{gc1}+N_2/(N_1+N_2)S_{gc2}$.} There are also three possibilities, which is similar to the fraction of active nodes. Fig.~\ref{fig5} shows the interaction between $S_{gc1}$ and $S_{gc2}$. In the case with two jumps, when a jump appears in $S_{gci}$, there is also a jump appearing in $S_{gcj}(i\neq j)$. So there are two discontinuous jumps in both $S_{gc1}$ and $S_{gc2}$. \textcolor{black}{We can discuss the emergence of the discontinuous jumps in a similar way as we discussed the functions $S_i(f)$ in Fig.~\ref{fig2} and Fig.~\ref{fig3} in the previous section.}

\section{\label{sec4}Conclusion}
We have studied the bootstrap percolation on complex network with bi-community structure, in which we observe either continuous appearance of the giant active component, or discontinuous hybrid transition of that. In contrast to the unstructured network, community structured networks may exhibit multiple discontinuous transitions for the fraction of the giant active component. We find that the discontinuous transition in one community triggers a simultaneous discontinuous jump in the other. The number of discontinuous transitions depends on the degree distributions of the two communities and their connections. In Erd\H{o}s-R\'{e}nyi networks we observe that if the inner-degrees are comparable or one of which is small, the system shows at most one jump; otherwise it undergoes multiple hybrid phase transitions.
Our results exhibit important properties of information spreading dynamics on real social networks. Moreover, the network system with communities is exactly the same with the system of network of networks\cite{GBSH12}. So, our result also can be generalized to this kind of systems.
\section{\label{sec5}Acknowledgement}
This work is supported by the NSFC(Grants No. 61203156) \textcolor{black}{and the Fundamental Research Funds for the Central Universities (Project No. 2682014RC17 and SWJTU12BR032}) and MOE Youth Fund Project of Humanities and Social Sciences (Project No. 11YJC840006). \textcolor{black}{Thanks for the valuable suggestions and comments the referees pointed out.}

\end{document}


\date{}
\maketitle

\section{Appendix A}
Randomly choose an edge from subnetwork $i$ to subnetwork $j(i,j=1,2)$. $Z_{i,j}$ denotes the probability that the node arrived at is active in the steady state. Similar to $Z_{11}$,
\begin{equation*}
Z_{22}=f_2+(1-f_2)\sum_{i+j\geq k}\frac{(j+1)P^2(i,j+1)}{\sum_{i\geq 0,j\geq 0}jP^2(i,j)}\sum_{\begin{subarray}{l} l+m=k\\
l\leq i,m\leq j\end{subarray}}^{i+j}\binom{i}{l}Z^l_{21}(1-Z_{21})^{i-l}\binom{j}{m}Z_{22}^m(1-Z_{22})^{j-m}
\end{equation*}

\begin{equation*}
Z_{12}=f_2+(1-f_2)\sum_{i+j\geq k}\frac{(i+1)P^2(i+1,j)}{\sum_{i\geq 0,j\geq 0}iP^2(i,j)}\sum_{\begin{subarray}{l}l+m=k\\
l\leq i,m\leq j\end{subarray}}^{i+j}\binom{i}{l}Z^l_{21}(1-Z_{21})^{i-l}\binom{j}{m}Z_{22}^m(1-Z_{22})^{j-m}
\end{equation*}

\begin{equation*}
Z_{21}=f_1+(1-f_1)\sum_{i+j\geq k}\frac{(j+1)P^1(i,j+1)}{\sum_{i\geq 0,j\geq 0}jP^1(i,j)}\sum_{\begin{subarray}{l}l+m=k\\
l\leq i,m\leq j\end{subarray}}^{i+j}\binom{i}{l}Z_{11}^l(1-Z_{11})^{i-l}\binom{j}{m}Z_{12}^m(1-Z_{12})^{j-m}
\end{equation*}

The equation for $S_2$ of a node in subnetwork 2 to be active at the steady state is similar to the equation of $S_1$:
\begin{equation*}
S_{2}=f_2+(1-f_2)\sum_{i+j\geq k}P^2(i,j)\sum_{\begin{subarray}{l} l+m=k\\
l\leq i,m\leq j\end{subarray}}^{i+j}\binom{i}{l}Z_{21}^l(1-Z_{21})^{i-l}\binom{j}{m}Z^m_{22}(1-Z_{22})^{j-m}
\end{equation*}

\section{Appendix B}
We show that in ER network $Z_{11}=Z_{21}=S_{1}$. Notice that \\ $\sum_{i\geq 0}\frac{k_{11}^i e^{-k_{11}}}{i!}=1$, we have

\begin{align*}
\frac{(i+1)P^1(i+1,j)}{\sum_{i\geq 0,j\geq 0}iP^1(i,j)}&=\frac{(i+1)\cdot\frac{k_{11}^{i+1} e^{-k_{11}}}{(i+1)!} \cdot\frac{k_{12}^j e^{-k_{12}}}{j!}}
{\sum_{i\geq 0,j\geq 0}i\cdot\frac{k_{11}^{i} e^{-k_{11}}}{i!}\cdot\frac{k_{12}^j e^{-k_{12}}}{j!}}\nonumber\\
&=\frac{\frac{k_{11}^{i+1}}{i!} e^{-k_{11}}\frac{k_{12}^j e^{-k_{12}}}{j!}}{\sum_{j\geq 0}(\sum_{i\geq 1}
\frac{k_{11}^{i-1} e^{-k_{11}}}{(i-1)!})k_{11}\frac{k_{12}^j e^{-k_{12}}}{j!}}=P^1(i,j).
\end{align*}

So $Z_{11}=S_{1}$. Similarly $Z_{21}=S_{1}$ and $Z_{22}=Z_{12}=S_{2}$.

Suppose a randomly selected node in subnetwork 1  has $r$ active neighbors. Let $P(r)$ be the corresponding probability. Suppose $l$ active neighbors are in subnetwork 1. Let $N_i(i=1,2)$ be the number of the nodes in each subnetwork respectively. Then the number of all neighbors of the node must be larger than $r$, and less than $N_1+N_2$. Let $m$ be the number of all neighbors and $i$ be the number of neighbors in subnetwork 1, then $l\leq i\leq m$ and $r\leq m\leq N_1+N_2$.
$$P(r)=\sum_{l=0}^r\sum_{m=r}^{N_1+N_2}\sum_{i=l}^mP^1(i,m-i)\binom{i}{l}Z_{11}^l(1-Z_{11})^{i-l}\binom{m-i}{r-l}Z_{12}^{r-l}(1-Z_{12})^{m-i-(r-l)},$$
and
$$S_{1}=f_1+(1-f_1)\sum_{r\geq k}P(r).$$
\begin{align}
P(r)&=\sum_{l=0}^r\sum_{m=r}^{N_1+N_2}\sum_{i=l}^m\binom{N_1}{i}
(p_{11})^i(1-p_{11})^{N_1-i}\binom{N_2}{m-i}p_{12}^{m-i}(1-p_{12})^{N_2-(m-i)}\nonumber\\
& \binom{i}{l}Z_{11}^l(1-Z_{11})^{i-l}\binom{m-i}{r-l}Z_{12}^{r-l}(1-Z_{12})^{m-i-(r-l)}.\nonumber
\end{align}
Here let $i=l+i'$, $m=r+m'$, $m-i=r-l+m'-i'$.
Notice that when $i=m$, $r=l$, we have $m'+r-l=m'$, then
\begin{align}
P(r)&=\sum_{l=0}^r\sum_{m'=0}^{N_1+N_2-r}\sum_{i'=0}^{m'}
(p_{11})^{l+i'}(1-p_{11})^{N_1-l-i'}\binom{N_1}{l+i'}\binom{l+i'}{l}Z_{11}^l(1-Z_{11})^{i'}\nonumber\\
& p_{12}^{r-l+m'-i'}(1-p_{12})^{N_2-(r-l)-(m'-i')}\binom{N_2}{m-i}\binom{m-i}{r-l}Z_{12}^{r-l}(1-Z_{12})^{m'-i'}\nonumber\\
&=\sum_{l=0}^rZ_{11}^l (p_{11})^l\binom{N_1}{l}Z_{12}^{r-l}(p_{12})^{r-l}\binom{N_2}{r-l}\sum_{m'=0}^{N_1+N_2-r}\sum_{i'=0}^{m'}
(p_{11})^{i'}(1-p_{11})^{N_1-l-i'}\nonumber\\
& \binom{N_1-l}{i'}(1-Z_{11})^{i'}p_{12}^{m'-i'}(1-p_{12})^{N_2-(r-l)-(m'-i')}\binom{N_2-(r-l)}{m'-i'}(1-Z_{12})^{m'-i'}.\nonumber
\end{align}
On the other hand,
$$p_{11}^k(1-p_{11})^{N_1-l-k}\binom{N_1-l}{k}\approx\frac{k_{11}^k e^{-k_{11}}}{k!}.$$
Therefore,
\begin{align}
P(r)&=\sum_{l=0}^rZ_{11}^l(p_{11})^l\binom{N_1}{l}Z_{12}^{r-l}(p_{12})^{r-l}\binom{N_2}{r-l}\nonumber\\
&\sum_{m'=0}^{N_1+N_2-r} \sum_{i'=0}^{m'} \biggl\{\frac{[k_{11}(1-Z_{11})]^{i'} e^{-k_{11}}}{i'!}\cdot\frac{[k_{12}(1-Z_{12})]^{m'-i'} e^{-k_{12}}}{(m'-i')!}\biggr\}\nonumber\\
&=\sum_{l=0}^rZ_{11}^l(p_{11})^l\binom{N_1}{l}Z_{12}^{r-l}(p_{12})^{r-l}\binom{N_2}{r-l}\nonumber\\
& \sum_{m'=0}^{N_1+N_2-r} \frac{[k_{11}(1-Z_{11})+k_{12}(1-Z_{12})]^{m'} e^{-[k_{11}(1-Z_{11})+k_{12}(1-Z_{12})]}}{m'!} e^{-(k_{11}Z_{11}+k_{12}Z_{12})}. \nonumber
\end{align}
Notice that the last sum is 1 in the infinite size limit, so
\begin{equation*}
P(r)=\sum_{l=0}^rZ_{11}^lp_{11}^l\binom{N_1}{l}Z_{12}^{r-l}p_{12}^{r-l}\binom{N_2}{r-l} e^{-(k_{11}Z_{11}+k_{12}Z_{12})}.
\end{equation*}
When $N$ is large enough, $l$ is relatively small, $\frac{N!}{N^l(N-l)!}\approx 1$, so
\begin{align}
P(r)&=\sum_{l=0}^rZ_{11}^lp_{11}^l\frac{N_1^l}{l!}Z_{12}^{r-l}p_{12}^{r-l}\frac{N_2^{r-l}}{(r-l)!} e^{-(k_{11}Z_{11}+k_{12}Z_{12})}\nonumber\\
&=\sum_{l=0}^r\frac{r!}{r!}\frac{(Z_{11}k_{11})^l}{l!}\frac{(Z_{12}k_{12})^{r-l}}{(r-l)!} e^{-(k_{11}Z_{11}+k_{12}Z_{12})}\nonumber\\
&=\frac{(Z_{11}k_{11}+Z_{12}k_{12})^r}{r!} e^{-(k_{11}Z_{11}+k_{12}Z_{12})}.\nonumber
\end{align}
So we can get the equation for $S_1$.

\section{Appendix C}
$X_{ij}$ is the probability that the node arrived at by following an arbitrarily chosen edge from community $i$ to community $j$ satisfying the conditions for $Z_{ij}$ and has at least one edge leading to an active subtree of infinite extent except the arbitrarily chosen edge. There are two possibilities: one is that the node arrived at is active at first and has at least one edge leading to an active subtree of infinite extent; the other one is that the node arrived at is inactive initially, but it has more than $k$ edges, except the arbitrarily chosen edge, leading to active nodes at equilibrium in community 1 or 2, and at least one of those edges leads to an active subtree of infinite extent.

The equation for $X_{11}$ is
\begin{align*}
X_{11}&=f_1\sum_{i+j\geq 1}\frac{(i+1)P^1(i+1,j)}{\sum_{i\geq 0,j\geq 0}iP^1(i,j)}\biggl[\sum_{\begin{subarray}{l}0\leq m\leq i\\0\leq n\leq j\\m+n\geq 1\end{subarray}}\binom{i}{m}X_{11}^m(1-X_{11})^{i-m}\binom{j}{n}X_{12}^n(1-X_{12})^{j-n}\biggr]\nonumber\\
&+(1-f_1)\sum_{i+j\geq k}\frac{(i+1)P^1(i+1,j)}{\sum_{i\geq 0,j\geq 0}iP^1(i,j)}\biggl\{\sum_{\begin{subarray}{l}k\leq m+n\leq i+j\\
0\leq m\leq i\\0\leq n\leq j\end{subarray}}\binom{i}{m}\binom{j}{n}\nonumber\\&
\biggl[\sum_{\begin{subarray}{l}1\leq s+t\leq m+n\\0\leq s\leq m\\0\leq t\leq n \end{subarray}}\binom{m}{s}\binom{n}{t}X_{11}^s(Z_{11}-X_{11})^{m-s} X_{12}^t(Z_{12}-X_{12})^{n-t}\nonumber\\&(1-Z_{11})^{i-m}(1-Z_{12})^{j-n}\biggr]\biggr\}.
\end{align*}

In the equation, $\sum_{i+j\geq 1}\frac{(i+1)P^1(i+1,j)}{\sum_{i\geq 0,j\geq 0}iP^1(i,j)}$ is the probability that the node we arrive at has $i$ other edges(except the one we arrived from) connecting to the nodes in community 1 and $j$ edges connecting to the nodes in community 2. $\sum_{\begin{subarray}{l}0\leq m\leq i\\0\leq n\leq j\\m+n\geq 1\end{subarray}}\binom{i}{m}X_{11}^m(1-X_{11})^{i-m}\binom{j}{n}X_{12}^n(1-X_{12})^{j-n}$ is the probability that in the $i+j$ edges there is at least one edge leading to an active subtree of infinite extent, and the edges leading to active subtree of infinite extent may connect to nodes in community 1 or 2, as the active subtree of infinite extent is in the whole network. That is, the two gccs in two communities are connected. $\sum_{\begin{subarray}{l}k\leq m+n\leq i+j\\
0\leq m\leq i\\0\leq n\leq j\end{subarray}}\binom{i}{m}\binom{j}{n}
\biggl[\sum_{\begin{subarray}{l}1\leq s+t\leq m+n\\0\leq s\leq m\\0\leq t\leq n \end{subarray}}\binom{m}{s}\binom{n}{t}X_{11}^s(Z_{11}-X_{11})^{m-s} X_{12}^t(Z_{12}-X_{12})^{n-t}(1-Z_{11})^{i-m}(1-Z_{12})^{j-n}\biggr]$ is the probability that the node we arrived at has more than $k$ edges, except the arbitrarily chosen edge, leading to active nodes at equilibrium in community 1 or 2, and at least one of those edges leads to an active subtree of infinite extent.

The equation for  $S_{gc2}$ that an arbitrarily chosen node in subnetwork 2 belongs to the giant active component is similar to the equation for $S_{gc1}$:
\begin{align*}
S_{gc2}&=f_2\sum_{i+j\geq 1}P^2(i,j)
\biggl[\sum_{\begin{subarray}{l}0\leq m\leq i\\
 0\leq n\leq j\\m+n\geq 1\end{subarray}}\binom{i}{m}X_{21}^m(1-X_{21})^{i-m}\binom{j}{n}X_{22}^n(1-X_{22})^{j-n}\biggr]\nonumber\\
&+(1-f_2)\sum_{i+j\geq k}P^2(i,j)\biggl\{\sum_{\begin{subarray}{l}k\leq m+n\leq i+j\\ 0\leq m\leq i\\0\leq n\leq j\end{subarray}}\binom{i}{m}\binom{j}{n}
\biggl[\sum_{\begin{subarray}{l}1\leq s+t\leq m+n\\0\leq s\leq m\\0\leq t\leq n \end{subarray}}\binom{m}{s}\binom{n}{t} \nonumber\\
& X_{21}^s(Z_{21}-X_{21})^{m-s}X_{22}^t(Z_{22}-X_{22})^{n-t}(1-Z_{21})^{i-m}(1-Z_{22})^{j-n}\biggr]\biggr\}.
\end{align*}

\section{Appendix D}
In ER network, the equation for $S_{gc1}$ that an arbitrarily chosen node in subnetwork 1 belongs to the giant active component is
\begin{align*}
S_{gc1}&=f_1\biggl\{\sum_{i+j\geq 1}P^1(i,j) \biggl[\sum_{\begin{subarray}{l}0\leq m\leq i\\ 0\leq n\leq j\\m+n\geq 1\end{subarray}}\binom{i}{m}\binom{j}{n}
S_{gc1}^m (1-S_{gc1})^{i-m} S_{gc2}^n (1-S_{gc2})^{j-n} \biggr]\biggr\}\nonumber\\
 &+(1-f_1)\sum_{i+j\geq k}P^1(i,j)\biggl\{\sum_{\begin{subarray}{l}k\leq m+n\leq i+j\\ 0\leq m\leq i\\0\leq n\leq j\end{subarray}}\binom{i}{m}\binom{j}{n}
 \biggl[\sum_{\begin{subarray}{l}1\leq s+t\leq m+n\\0\leq s\leq m\\0\leq t\leq n \end{subarray}}\binom{m}{s}\binom{n}{t}\\& S_{gc1}^s(S_{1}-S_{gc1})^{m-s} S_{gc2}^t(S_{2}-S_{gc2})^{n-t}(1-S_{1})^{i-m}(1-S_{2})^{j-n}\biggr]\biggr\}.
\end{align*}

The equation for $S_{gc2}$ that an arbitrarily chosen node in subnetwork 2 belongs to the giant active component is similar to the equation for $S_{gc1}$:
\begin{align*}
S_{gc2}&=f_2\biggl\{\sum_{i+j\geq 1}P^2(i,j)
\biggl[\sum_{\begin{subarray}{l}0\leq m\leq i\\
 0\leq n\leq j\\m+n\geq 1\end{subarray}}\binom{i}{m}S_{gc1}^m(1-S_{gc1})^{i-m}\binom{j}{n}S_{gc2}^n(1-S_{gc2})^{j-n}\biggr]\biggr\}\nonumber\\
&+(1-f_2)\sum_{i+j\geq k}P^2(i,j)\biggl\{\sum_{\begin{subarray}{l}k\leq m+n\leq i+j\\ 0\leq m\leq i\\0\leq n\leq j\end{subarray}}\binom{i}{m}\binom{j}{n}
\biggl[\sum_{\begin{subarray}{l}1\leq s+t\leq m+n\\0\leq s\leq m\\0\leq t\leq n \end{subarray}}\binom{m}{s}\binom{n}{t} \nonumber\\
& S_{gc1}^s(S_{1}-S_{gc1})^{m-s}S_{gc2}^t(S_{2}-S_{gc2})^{n-t}(1-S_{1})^{i-m}(1-S_{2})^{j-n}\biggr]\biggr\}.
\end{align*}

\section{Appendix E}

\begin{figure}
\centering
\includegraphics[scale=0.6]{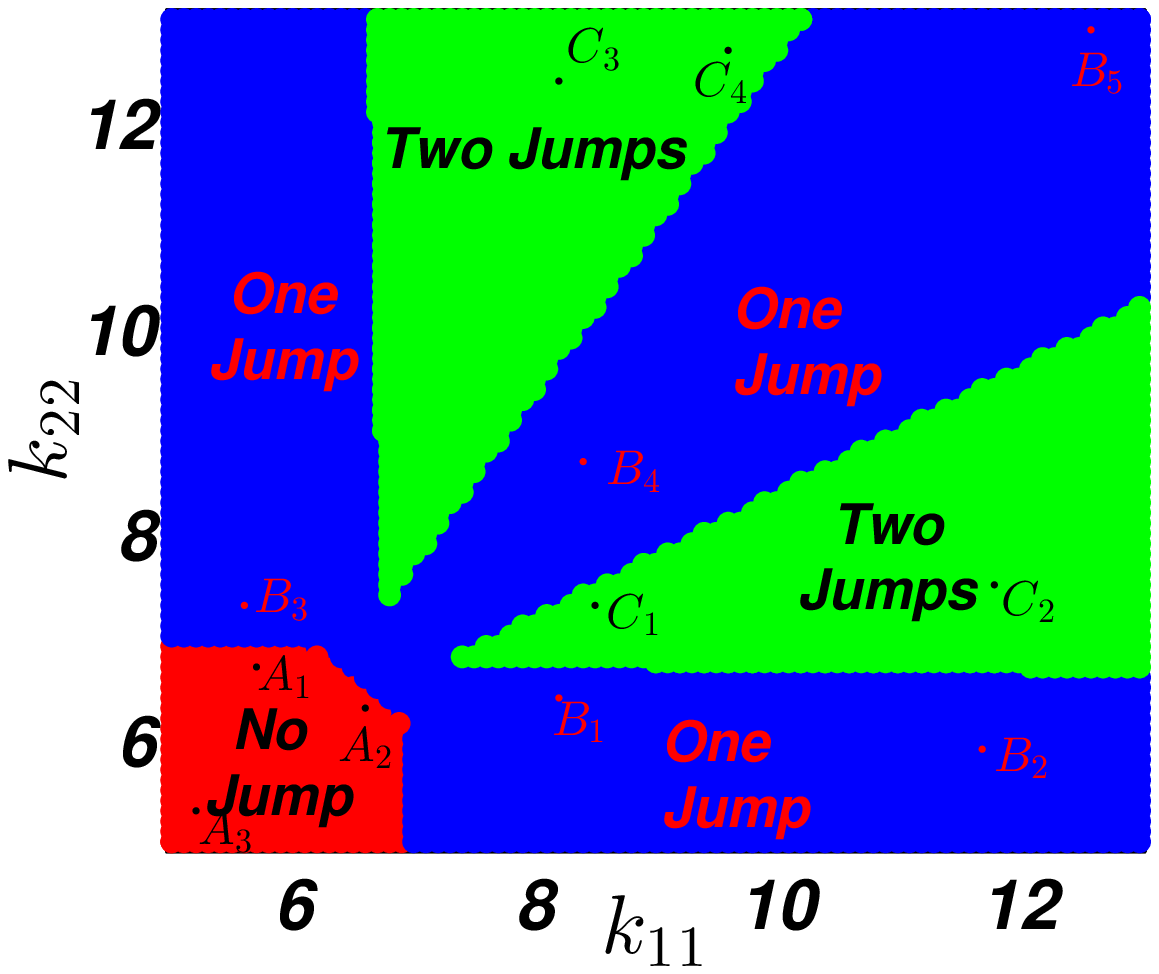}
\caption{The diagram of the phase transitions by theoretical calculation}
\label{newPhaseDiagram}
\end{figure}

\begin{figure}
\centering
\includegraphics[scale=0.5]{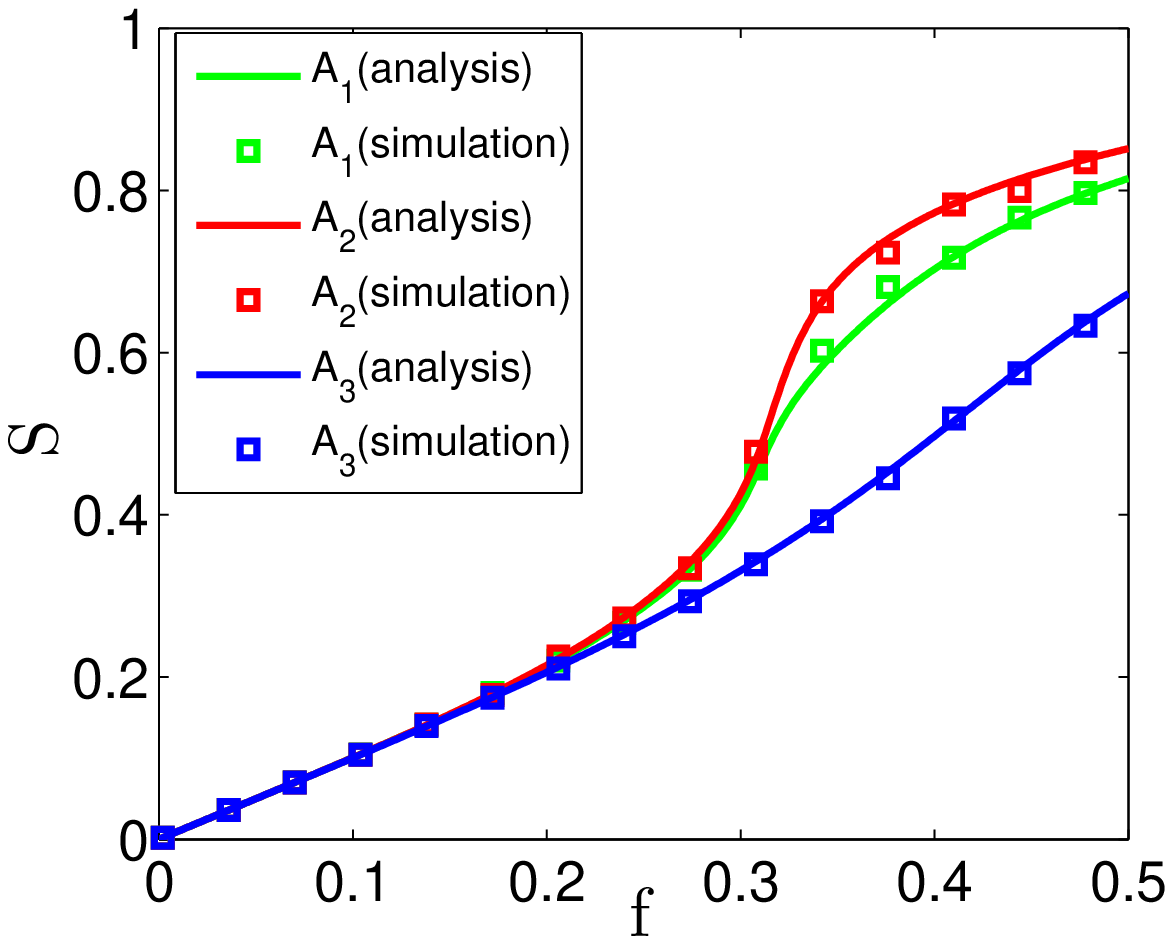}\includegraphics[scale=0.5]{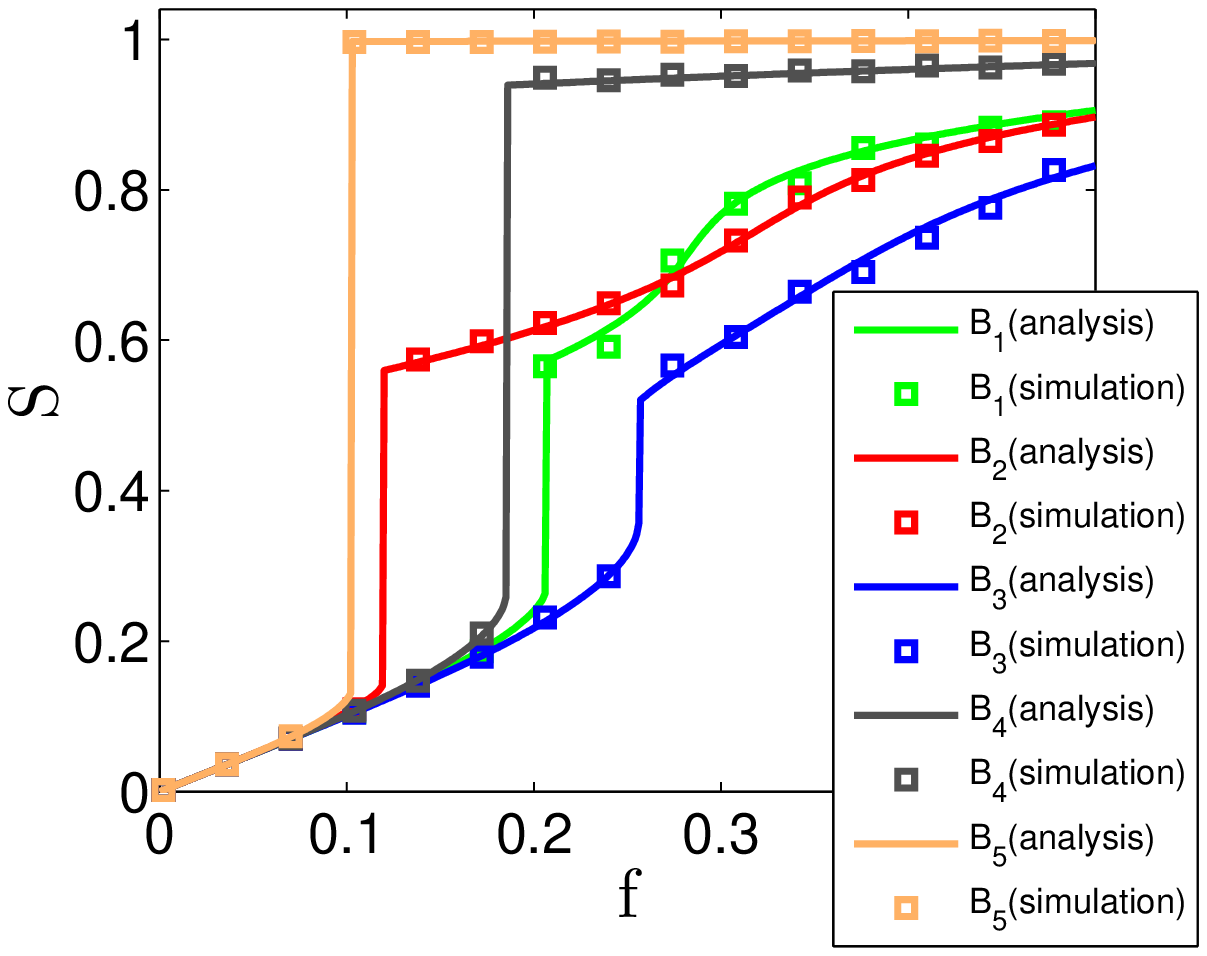}
\includegraphics[scale=0.5]{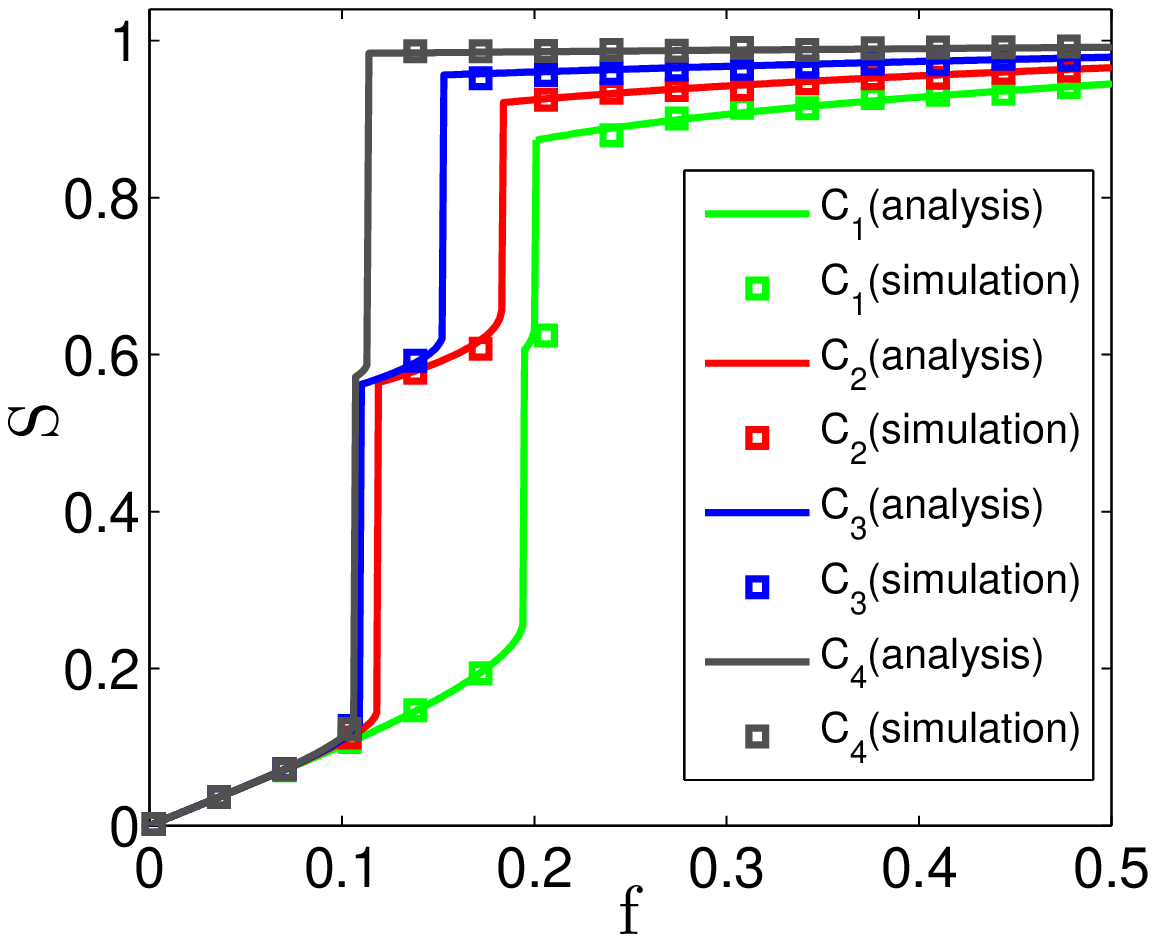}
\caption{Comparison of theoretical results and simulations in terms of the fraction $S$ of the active nodes in the entire network.  Symbols are simulation data. Solid lines are theoretical solutions. We assume $S=\frac{S_1+S_2}{2}$, and $f_1=f_2=f$. Here we set $k_{12}=0.5$, $k_{21}=0.5$, and $k=5$.}
\label{comparison}
\end{figure}

Fig.~\ref{newPhaseDiagram} is the phase diagram we showed in our manuscript.  We randomly chose points $A_i, B_i, C_i$ in different regions of the phase diagram as shown in the figure, and showed the active fraction $S=(S_1+S_2)/2$ in the whole network as a function of the initial active fraction  $f_1=f_2=f$ for different groups of values of $k_{ij}$ in an ER network, both theoretically and by simulations in Fig.~\ref{comparison}.
It can be seen that theoretical prediction is consistent with the simulation results.

\begin{figure}
\centering
\includegraphics[scale=0.33]{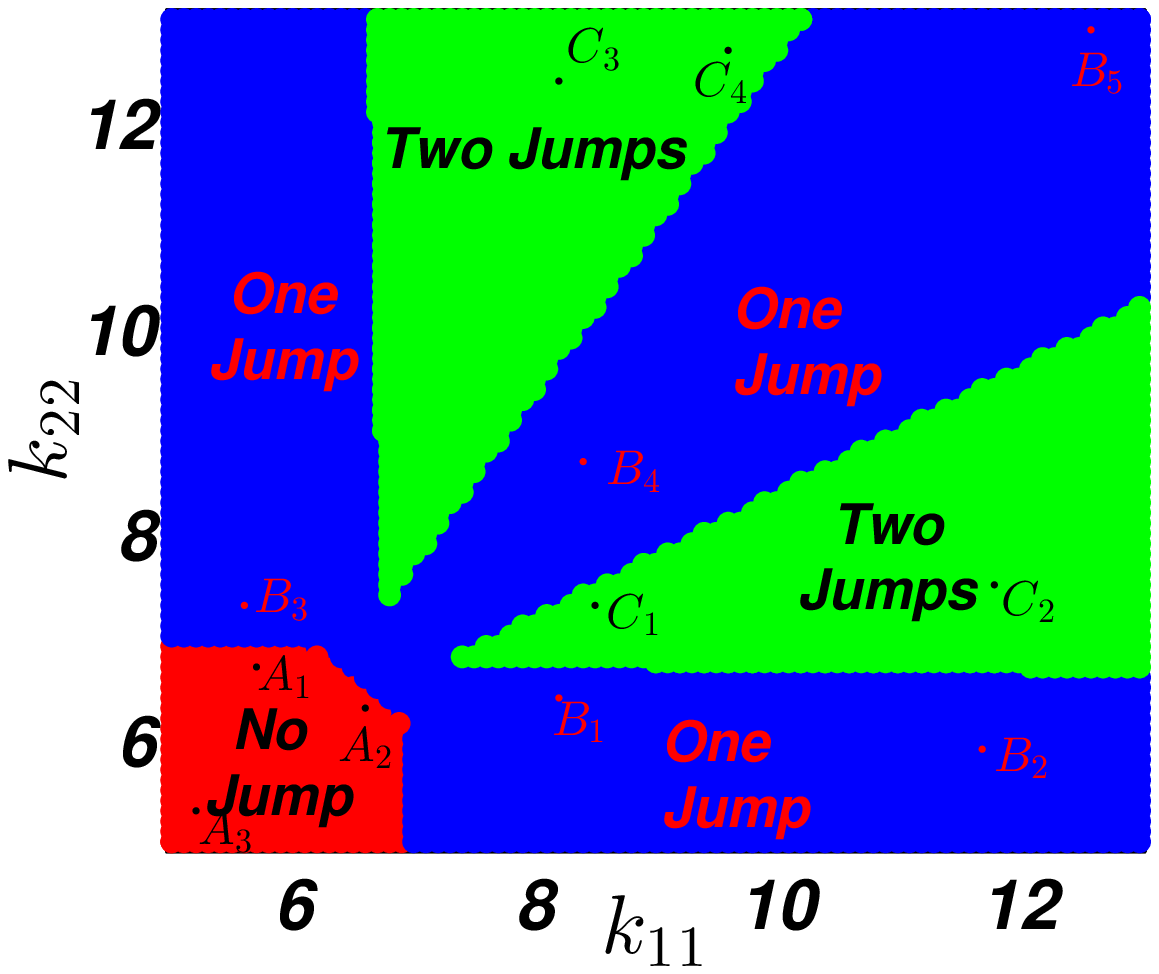}\includegraphics[scale=0.375]{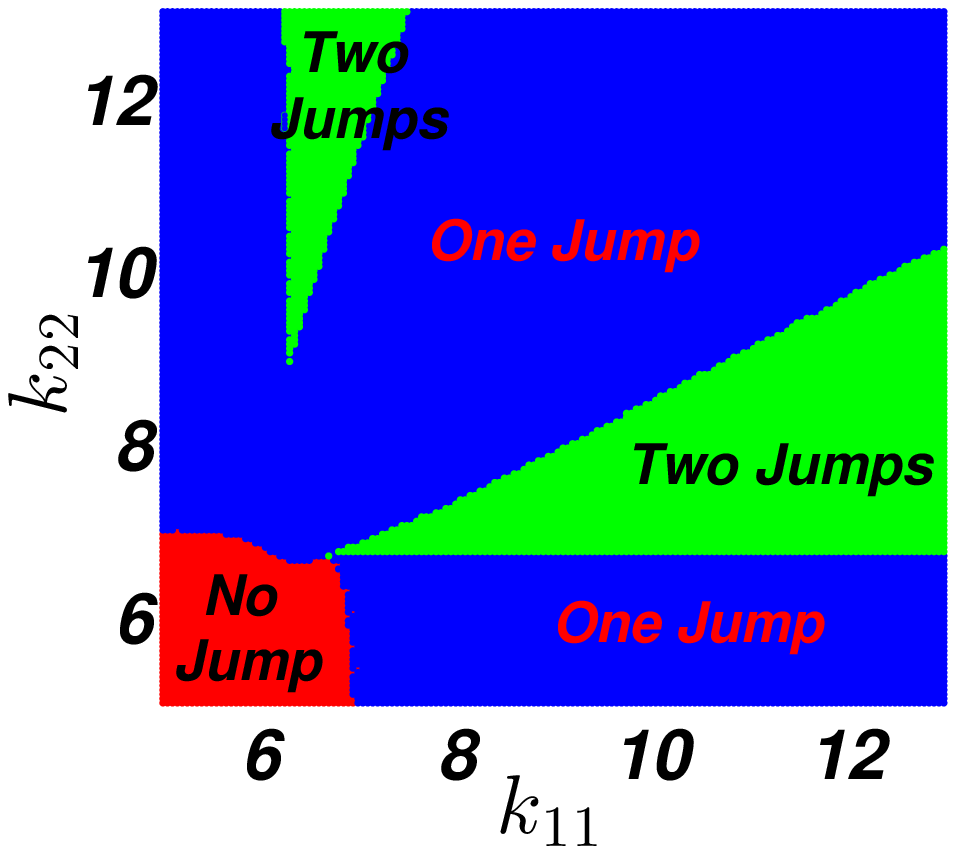}\includegraphics[scale=0.375]{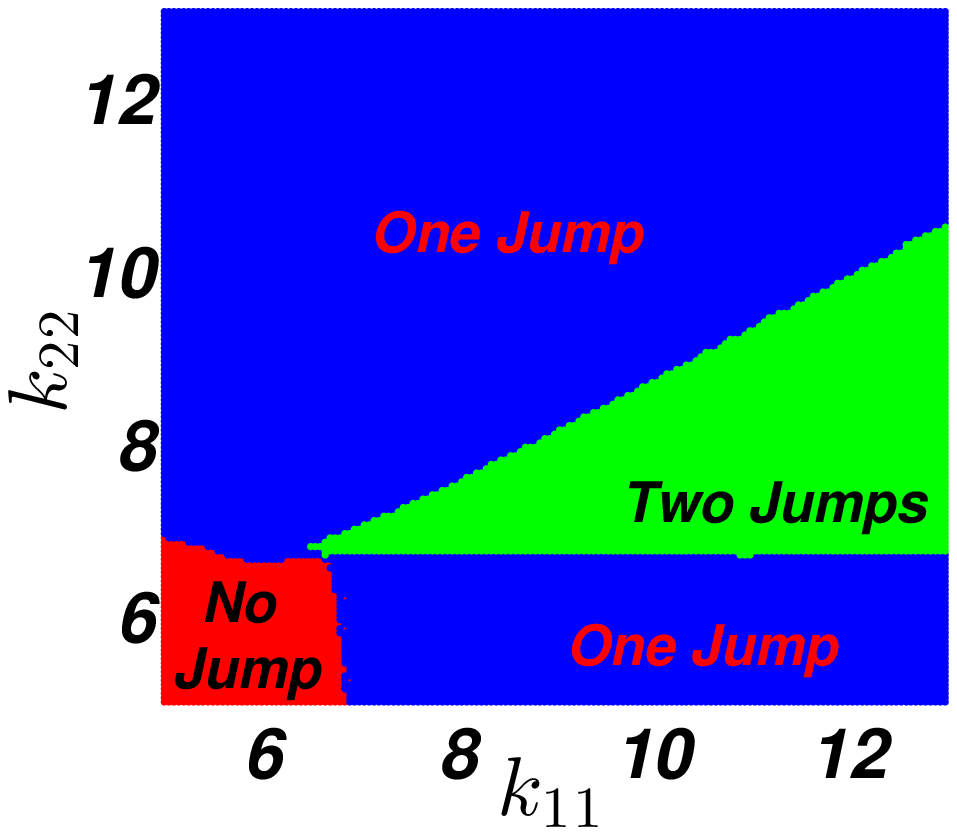}
\caption{The phase diagram when $N_1=N_2$, $2N_1=N_2$, $3N_1=N_2$ respectively.}
\label{different diagrams}
\end{figure}

When $N_1\neq N_2$, that is, when $k_{12}\neq k_{21}$, the phase diagram is different.
Since the functions $S_i$ depend on  the values of $k_{12}, k_{21}, k , f_i$,
the phase diagram also depends on the inner-degrees and out-degrees of the network $k_{ij}$.
Notice that if $N_1$ is not equal to $N_2$, $k_{12}$ will be not equal to $k_{21}$ either, then it will affect the phase diagram, so the diagram will be different from Fig.~1 in our manuscript. We show the phase diagram when $N_1\neq N_2$, that is, $k_{12}\neq k_{21}$ in Fig.~\ref{different diagrams} here. We suppose $k_{21}=0.5, k=5, f_1=f_2=f$, and let $k_{12}=0.5$, $k_{12}=1$, $k_{12}=1.5$ respectively, that is, when $N_1=N_2$, $2N_1=N_2$, $3N_1=N_2$ respectively.

\section{Appendix F}
\begin{figure}
\centering
\includegraphics[scale=0.6]{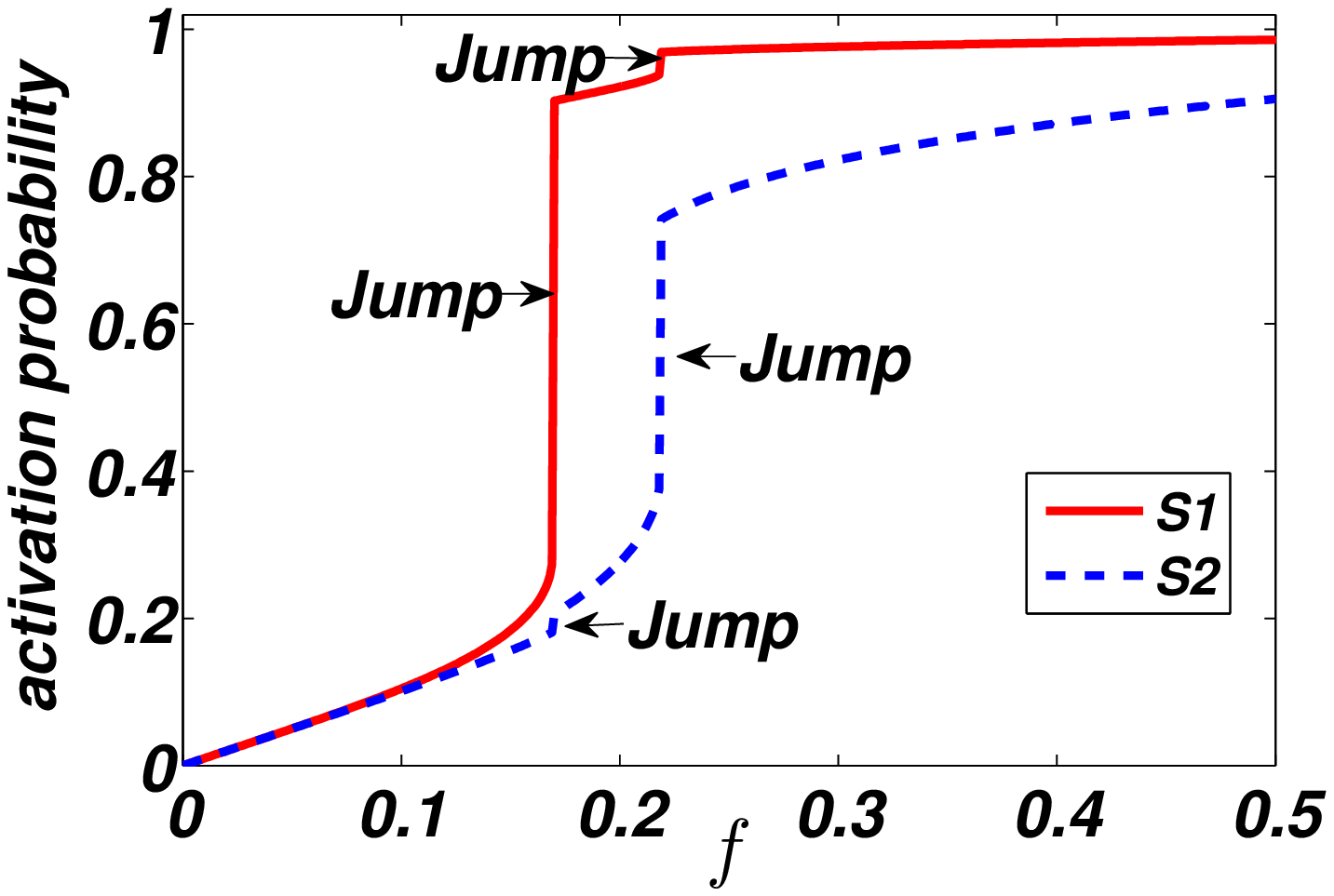}
\caption{(Color online) The interaction between $S_1$ and $S_2$. There are two discontinuous jumps in both $S_1$ and $S_2$. When a jump appears in $S_i$, there is also a jump in $S_j(i\neq j)$. Here $k_{11}=8$, $k_{22}=7$, $k_{12}=2.4$, $k_{21}=0.5$, $k=5$.}
\label{fig2}
\end{figure}

 As functions of $f$, $S_1$ and $S_2$  interact with each other.  We can see in Fig.~\ref{fig2} that when a jump appears in $S_i$, there is also a jump in $S_j(i\neq j)$.
